# Tailoring Bulk Photovoltaic Effects in Magnetic Sliding Ferroelectric Materials

Chunmei Zhang[1], Ping Guo[1], Jian Zhou[2,*]

[1]School of Physics, Northwest University, Xi'an 710069, China

[2]Center for Alloy Innovation and Design, State Key Laboratory for Mechanical Behavior of Materials, Xi'an Jiaotong University, Xi'an 710049, China

[*]Email: jianzhou@xjtu.edu.cn

## Abstract

The bulk photovoltaic effect that is intimately associated with crystalline symmetry has been extensively studied in various nonmagnetic materials, especially ferroelectrics with a switchable electric polarization. In order to further engineer the symmetry, one could resort to spin-polarized systems possessing an extra magnetic degree of freedom. Here, we investigate the bulk photovoltaic effect in two-dimensional magnetic sliding ferroelectric (MSFE) systems, illustrated in $VSe_2$, $FeCl_2$, and $CrI_3$ bilayers. The transition metal elements in these systems exhibit intrinsic spin polarization, and the stacking mismatch between the two layers produce a finite out-of-plane electric dipole. Through symmetry analyses and first-principles calculations, we show that photoinduced in-plane bulk photovoltaic current can be effectively tuned by their magnetic order and the out-of-plane dipole moment. The underlying mechanism is elucidated from the quantum metric dipole distribution in the reciprocal space. The ease of the fabrication and manipulation of MSFEs guarantee practical optoelectronic applications.

TOC figure

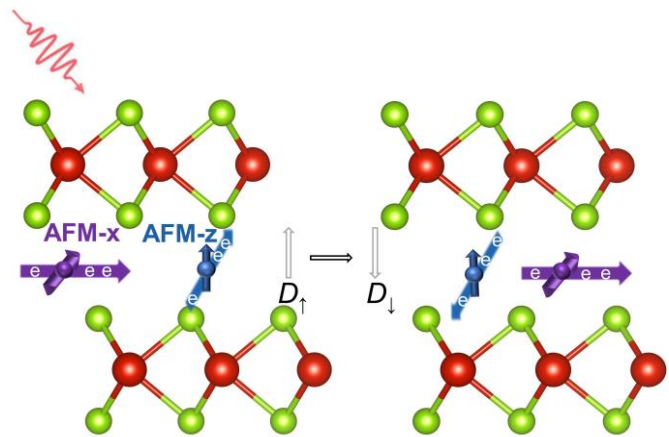

Bulk photovoltaic (BPV) effect, a second order nonlinear optical (NLO) response, could produce steady-state electric current under homogeneous optical illumination.[1-3] According to theoretical estimations, it may host the potential to overpass the well-known Shockley-Queisser limit in conventional solar cells, yet one does not require complicated *p-n* heterojunction fabrications for the photo-electric conversion.[4] Fundamentally, the BPV process serves as an efficient tool for detecting the electronic topology.[5] Extensive studies on BPV effect in time-reversal ($\mathcal{T}$) invariant systems have been conducted, including their shift current,[6-8] injection current,[9] nonlinear Hall current generations,[10] etc.[11] They are rooted in various geometric phases of the electronic wavefunctions, namely, shift vector, Berry connection, and Berry curvature dipole, respectively. These features assign that the BPV currents only emerge in centrosymmetry ($\mathcal{P}$) broken systems, e.g., ferroelectric materials.[12-15]

Two-dimensional (2D) materials are attractive for their good optical accessibility, compared with the conventional 3D bulk systems, as their ultrahigh surface-to-volume ratio and marginal light dispersion vertical to the atomic plane. As such, the phase-matching condition in NLO processes (e.g., second harmonic generation) is naturally satisfied. Over the past decade, several 2D ferroelectric materials have been theoretically predicted and then realized in subsequent experiments, such as group-IV monochalcogenides,[16] $In_2Se_3$ monolayer,[17] $CuInP_2S_6$ ultrathin flakes,[18] $BiFeO_3$ in one unit-cell thick,[19] *d*1T-$MoTe_2$ monolayer,[20] etc. Nonetheless, the scarcity of intrinsic 2D ferroelectric systems limits their practical usage, and other routes to introducing electric dipoles need further exploration. Recently, Wu *et al.* proposed the concept of interfacial ferroelectricity in 2D van der Waals (vdW) materials, which could generate sizable out-of-plane dipole moments ($D_z$) in multilayer forms that are non-ferroelectric in their monolayer counterparts.[21-22] The flipping of $D_z$ (between $D_\uparrow$ and $D_\downarrow$) can be done via a short distance shuffling or sliding between neighboring vdW layers. Hence, they are referred to as sliding ferroelectric (SFE) materials.[23-26] Owing to the low energy barrier (~0.1 $\mu J/cm^2$) separating the $D_\uparrow$ and $D_\downarrow$ states, the flipping could occur in an ultrafast kinetics with high contrast. This motivates various experimental investigations after the theoretical predictions, and several SFEs have been demonstrated in *h*-BN, H and T′ phases of transition metal dichalcogenides, etc.[20, 27]

The out-of-plane dipole moment in SFE bilayers constraints that only nonlinear Hall current can be switched under $D_z$ flipping, while both in-plane shift and injection currents maintain their direction and magnitude.[28-29] In order to further toggle the BPV currents in SFEs, we propose that breaking $\mathcal{T}$-symmetry is an efficient way which adds another tuning parameter, namely, magnetic order (spin polarization direction). The interplay between different ferroic orders, such as ferromagnetism and ferroelectricity, has led to tremendous exotic physical properties with novel applications.[30-31] Here both spin and electric polarizations offer large space to engineer the symmetry, thus the light-matter interaction in the magnetic ferroelectric materials would provide additional manipulation opportunities for the BPV current generations.[32]

Recently, it has been shown that in magnetic systems the normal BPV (shift and injection) currents could exhibit magnetic counterparts, which vanish in nonmagnetic materials.[33] These magnetic cousins are dubbed magnetic shift current and magnetic injection current (MIC), arising from the inequivalent distribution of the Kramers pair states at **k** and −**k**.[34] Several NLO investigations in magnetic systems have been carried out, and special care has been given to $\mathcal{PT}$-symmetry, e.g., antiferromagnetic (AFM) bilayer $CrI_3$ and $MnBi_2Te_4$.[32,34-36] In such cases, the normal shift current (NSC) and normal injection current diminish. In this Letter, we consider magnetic SFE (referred to as MSFE) systems, in which all the $\mathcal{P}$, $\mathcal{T}$, and $\mathcal{PT}$ symmetries are broken. Thus, they host both normal and magnetic BPV currents simultaneously. The tunability of magnetic order and electric dipole provide a vast space for the BPV current manipulation.

We use $H$-$VSe_2$ and $T$-$FeCl_2$ bilayers to illustrate this concept, and perform first-principles density functional theory (DFT) calculations to evaluate the BPV photoconductivity. Note that in the discovery of both low-dimensional magnetic materials and SFE concepts, theoretical approaches (mainly according to DFT calculations) have shown its powerful predicting ability with good accuracy. These prior calculations have evoked several subsequent experimental investigations, e.g., monolayer $VSe_2$,[37-38] $CrI_3$,[39-41] $Fe_3GeTe_2$,[42-43] CrSBr,[44-45] $MnSe_2$,[46-47] etc.[22-23,28,48-51] According to previous theoretical works, both $H$-$VSe_2$[52-53] and $T$-$FeCl_2$[54] can be physically exfoliated from their bulk counterparts. They belong to hexagonal lattice and exhibit good chemical and thermodynamic stability. Their optimal bilayer stacking induces finite interfacial $D_z$. The unfilled $d$ orbitals in the transition metals produce intrinsic magnetic polarizations, giving intralayer ferromagnetic (FM) and interlayer AFM configuration. These ensure that their bilayers belong to MSFE materials, where the magnetic order vector (**L** = **M**$_1$ − **M**$_2$, **M** being magnetic vector in each layer) and electric polarization $D_z$ can be efficiently controlled and modulated. We show that under linearly polarized light (LPL) irradiation, the MIC generation depends on both **L** and $D_z$. On the other hand, the LPL induced NSC remains in its direction regardless of **L** and $D_z$. Since the flow directions of NSC and MIC under certain magnetic cases are vertical to each other, these photocurrents can be individually detected and controlled without strong entanglement.

*Geometric and electronic properties of MSFEs.* All monolayers $H$-$VSe_2$[55], $T$-$FeCl_2$[56] and $CrI_3$ belong to hexagonal lattice. Similar as that in the monolayer $MoS_2$,[51] there are several SFE stacking configurations. In the main text, we will mainly focus on the $H$-$VSe_2$, while leaving results and discussions for the $T$-$FeCl_2$ and $CrI_3$ in Supporting Information (SI). In Figure 1a, we plot the atomic structures of a high symmetric VSe$_2$ bilayer, where the upper layer is eclipsed over the lower one. This is denoted as the intermediate (IM) state, as it is energetically unstable and would spontaneously slide into two energetically degenerate patterns, namely, $D_\uparrow$ (Figure 1b) and $D_\downarrow$ (Figure 1c). Our calculation shows their dipole moment to be ±0.078 μC/cm$^2$, larger than the experimentally observed values of 0.032 μC/cm$^2$ in bilayer WTe$_2$.[57] The $D_\uparrow$ and $D_\downarrow$ states are subject to an in-plane sliding $t\left(\frac{1}{3},\frac{2}{3},0\right)$ and $t\left(\frac{2}{3},\frac{1}{3},0\right)$ (direct coordinates relative to lattice) from IM, respectively, and the switching entails an energy barrier of 64 meV per formula unit (f.u.), comparable to that of the In$_2$Se$_3$ monolayer (66 meV/f.u.).[58]

One notes that each Se favors a formal −2 reduction state, so that the V atom is in its +4 oxidation state. Therefore, each V leaves one unpaired electron, carrying ~1 μ$_B$ local magnetic moment. As elucidated previously,[38,59] the monolayer VSe$_2$ exhibits (intralayer) FM semiconducting ground state, with the valence band maximum (VBM) and conduction band minimum (CBM) locating at $K$ and $K'$ in the first Brillouin zone (BZ). When two VSe$_2$ layers are stacked, the interlayer coupling prefers AFM configuration, which is energetically lower than the FM interlayer state by 3 meV per unit cell. Our calculation results are well consistent with previous works.[60] We tabulate these results in Table S1. In this regard, the VSe$_2$ bilayer conceives both electric and magnetic polarizations simultaneously, making it a type-I multiferroic system.

The calculated band structures in different patterns are plotted in Figures 1d−1f. We assign the magnetic order **L** along $z$ (denoted as **L** ∥ **ẑ**). One sees that all of them exhibit an indirect bandgap semiconducting feature with both valence and conduction bands near the Fermi level mainly contributed by the V-$d$ orbitals. The VBM of the IM state at the $K$ and $K'$ valleys are energetically degenerate and contributed by the upper (spin up) and lower (spin down) layer, respectively (Figure 1d). This is because that the two layers can be mapped through a mirror reflection $\mathcal{M}_z$ multiplying $\mathcal{T}$, which gives eigenenergy $E(\mathbf{k}, \text{spin up}) = E(-\mathbf{k}, \text{spin down})$. Under finite $D_z$, the built-in electric potential lifts such degeneracy ($\mathcal{M}_z$-broken). According to our calculations, a large valley polarization of ~70 meV can be observed in the two MSFE states (Figures 1e and 1f). Compared with experimentally observed valley splitting of tens of meV in SiO$_2$/Si(100)/SiO$_2$ quantum well[61-62] and 2.5 meV under the magnetic field of 1 Tesla in WSe$_2$ monolayer,[63] such valley polarization is significant enough to be observed.

One can understand such valley polarization in the **L** ∥ **ẑ** VSe$_2$ MSFE using a simplified **k·p** model,

$$H = H_0 + H'_D + H'_{SOC} = H_0 + \lambda_z D_z \sigma_z + \lambda_{SOC}(3k_x^2 - k_y^2)k_y \sigma_z. \qquad (1)$$

Here, $H_0$ contains crystal field and magnetic exchange effect in the system, which is essentially the Hamiltonian of the IM state without SOC interactions. $H'_D$ is the electronic splitting induced by $D_z$, and the $H'_{SOC}$ is the SOC triggered spin splitting up to cubic **k**-terms in the $\mathcal{C}_{3v}$ system.[64] The wrapping term is included only because the Rashba SOC is associated with the in-plane spin polarization. The $\sigma_z$ is the $z$-component Pauli matrix for spin. $\lambda_z$ and $\lambda_{SOC}$ refer to the $D_z$-spin coupling strength and the intrinsic SOC strength, respectively.

The $D_z$ serves as an effective magnetic field that couples with spin and lifts the spin degeneracy, i.e., the reversal of $D_z$ flips spin up and spin down states (Figure S1). One also notes that $H'_{SOC}$ splits spin degeneracy at the $K$ and $K'$ valley. Thus, the synergistic effects of $D_z$ and SOC result in the valley polarization of the $D_\uparrow$ and $D_\downarrow$ bilayer VSe$_2$ (**L** ∥ $\hat{\mathbf{z}}$), and the magnitude scales with $\lambda_z$ and $\lambda_{SOC}$ (Figure S1). Remarkably, the dipole moment induces layer-imbalanced wavefunction distributions at any generic momenta pair (**k** and −**k**), which play a vital role in the injection photocurrent generation.

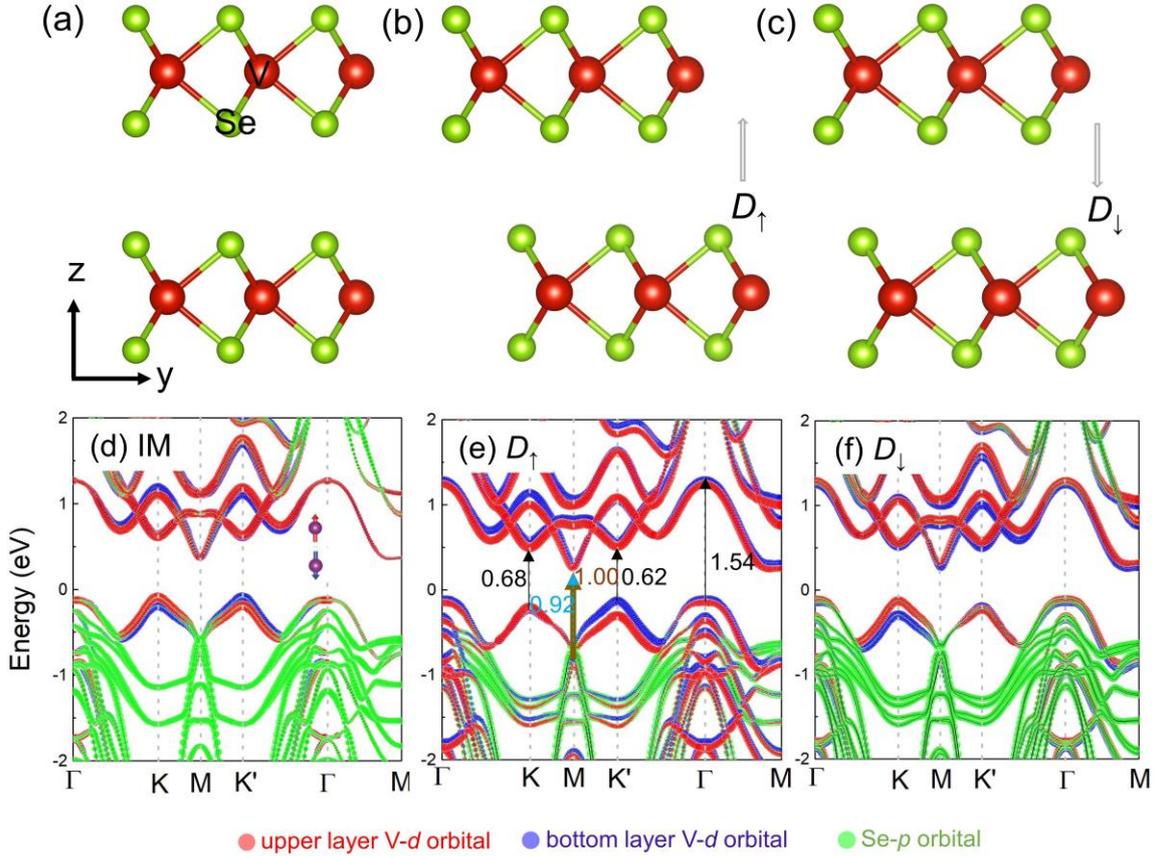

**Figure 1.** Atomic geometry of the (a) IM, (b) $D_\uparrow$, and (c) $D_\downarrow$ stacking patterns of bilayer VSe$_2$. (d)−(f) Calculated band structures for the IM, $D_\uparrow$, and $D_\downarrow$ states (**L** ∥ $\hat{\mathbf{z}}$). The energy is relative to the Fermi level. Values in panel (e) measure the direct bandgaps (in eV) at different **k** points.

*Symmetry considerations of photoconductivity.* We show that the symmetry of MSFE VSe$_2$ bilayer is very sensitive to **L** and $D_z$, which could switch the BPV photocurrents. Particularly, according to previous works,[34] LPL would induce both NSC and MIC generations. Note that for $\mathcal{T}$-symmetric systems, MIC vanishes, while NSC is symmetrically forbidden for $\mathcal{PT}$-systems. Both of them may exist in general cases, which is the situation in the MSFEs as studied here. Hence, we will consider both NSC and MIC generations, which carry geometric phases of electric wavefunctions, namely, shift vector and quantum metric dipoles (see Methods section).[10, 65-66]

The NSC and MIC photocurrent densities are[34]

$$J_{\text{NSC}}^a = \sigma^{abb}(0;\omega,-\omega)E_b(\omega)E_b(-\omega), \tag{2}$$

$$J_{\text{MIC}}^a = \eta^{abb}(0;\omega,-\omega)E_b(\omega)E_b(-\omega). \tag{3}$$

Here, the $\sigma^{abb}$ and $\eta^{abb}$ are the photoconductivity for NSC and MIC, respectively. $E$ is the alternating electric field (with angular frequency $\omega$), and $a$ ($b$) refers to Cartesian coordinate in the *xy* plane. Even though the higher order photo-responses would present as well. We note that when the light intensity is not significant, the photocurrent would reduce as the nonlinear order increases, according to the Kubo perturbation theory.[5, 67] Also, one notes that the static photocurrent does not exist for the third order photo-responses under a monochromatic light irradiation. In this regard, we only focus on the second order BPV effect in the current work.

According to the atomic coordination (ignoring spin polarization), the IM pattern belongs to the crystalline layer group $P\bar{6}m2$, which includes two mirror reflections with their normal direction along *x* ($\mathcal{M}_x$) and *z* ($\mathcal{M}_z$). The $D_\uparrow$ and $D_\downarrow$ states break $\mathcal{M}_z$, belonging to the layer group of $P3m1$. When we include spin polarization effect, the $\mathcal{M}_x$ becomes $\mathcal{M}_x\mathcal{T}$ for the **L** ∥ **ẑ**, since spin transforms as a pseudovector. For the IM state, the $\mathcal{M}_z$ changes into $\mathcal{M}_z\mathcal{T}$. However, if the magnetic moment on V switches its direction, such symmetry constraints would alter. For example, in the **L** ∥ **x̂** case, one has $\mathcal{M}_x$ reflection for all the three sliding structures, while the IM state contains $\mathcal{M}_z$. The **L** ∥ **ŷ** configuration is different, which needs multiplication of $\mathcal{T}$ onto $\mathcal{M}_x$ (for all three patterns). We list these symmetries in Table 1.

Next, we move to the symmetry constraints for NSC and MIC photoconductivities under LPL. According to Eq. (5) (see Methods), the NSC is defined as the overall contribution of shift vectors weighted by the absorption rate in the whole BZ.[68] As the absorption rate is always positive, the direction of the $\sigma^{abb}$ is scaled by the shift vector $R_{mn}^a$,[69] which transforms as a polar vector, giving that $\mathcal{M}_x R_{nm}^x(k_x, k_y) = -R_{nm}^x(-k_x, k_y)$ and is immune to $\mathcal{T}$. Hence, the $\sigma^{xxx}$ and $\sigma^{xyy}$ are symmetrically forbidden, leaving only $\sigma^{yxx}$ and $\sigma^{yyy}$ to be finite.

To check whether the direction of the NSC could be tuned by the dipole moment $D_z$, we further analyze the structure relationship between $D_\uparrow$ and $D_\downarrow$. Ignoring the spin order, the $D_\uparrow$ and $D_\downarrow$ spatial patterns are related to each other by $\mathcal{M}_z D_\uparrow = D_\downarrow$. The direction of the $\sigma^{abb}$ is controlled by the in-plane shift vector $R_{mn}^a$,[69] which does not flip sign upon the out-of-plane FE transition, $\mathcal{M}_z R_{mn}^a = R_{mn}^a$. Thus, the NSC keeps its direction regardless of $D_z$ and **L**.

As for the MIC generation, we show that **L** strongly affects the symmetry assignments. According to the Kubo perturbation theory and previous works,[34] the MIC arises from an excitation from bands $m$ to $n$ with the quantum metric tensor $g_{nm}^{bb}$, weighted by the asymmetric group velocity difference $\Delta_{nm}^a$ at $\pm\mathbf{k}$ (see Methods). The symmetry transformation gives $\mathcal{T}\Delta^a(\mathbf{k}) = -\Delta^a(-\mathbf{k})$ and $\mathcal{M}_x \Delta^a(\mathbf{k}) = (-1)^{\delta_{xa}} \Delta^a(\tilde{\mathbf{k}})$, where $\tilde{\mathbf{k}} = (-k_x, k_y)$.[70] The quantum metric $g_{nm}^{bb}$ transforms as $\mathcal{T} g_{nm}^{bb}(\mathbf{k}) = g_{nm}^{bb}(-\mathbf{k})$ and $\mathcal{M}_x g_{nm}^{bb}(\mathbf{k}) = g_{nm}^{bb}(\tilde{\mathbf{k}})$. Therefore, under ($x$ or $y$-polarized) LPL, for the $D_\uparrow$ and $D_\downarrow$ states, the MIC always transports along the $y$ direction for **L** ∥ $\hat{\mathbf{x}}$, while it flows along the $x$ direction for **L** ∥ $\hat{\mathbf{y}}$ and **L** ∥ $\hat{\mathbf{z}}$ situations.

For the **L** ∥ $\hat{\mathbf{x}}$ and **L** ∥ $\hat{\mathbf{y}}$ IM states, the mirror-$z$ reflection assigns constraints on the velocity difference as $\mathcal{M}_z \Delta^a(\mathbf{k}) = \Delta^a(\mathbf{k})$. Thus, they give similar MIC directions with those in both $D_\uparrow$ and $D_\downarrow$ (flowing along $x$ and $y$, respectively). When it comes to the **L** ∥ $\hat{\mathbf{z}}$ IM state, the symmetry constraint becomes $\mathcal{M}_z \mathcal{T} \Delta^a(\mathbf{k}) = -\Delta^a(-\mathbf{k})$, which diminishes both $\eta^{xxx}$ and $\eta^{xyy}$. Thus, the existence of $\mathcal{M}_z \mathcal{T}$ results in symmetric distribution and opposite sign of the $\Delta_{nm}^x g_{nm}^{xx}$ at $\mathbf{k}$ and $-\mathbf{k}$, so that $J_{\text{MIC}}^{xbb}$ is strictly to be zero. Accordingly, for the $D_\uparrow$ and $D_\downarrow$ states, the removal of $\mathcal{M}_z \mathcal{T}$ gives non-vanishing $J_{\text{MIC}}^{xbb}$.[32] These symmetric results for the MIC are listed in Table 1. In this table, the allowed (nonzero) MIC are given under each operation. In addition, the presence or absence of the MIC can also be estimated from the asymmetric bands at $\mathbf{k}$ and $-\mathbf{k}$, as shown in Figure S2.

**Table 1.** Magnetization orientation dependent symmetry operations and the allowed MIC for the IM, $D_\uparrow$, and $D_\downarrow$ states. We only list the AFM interlayer configuration results, and the symmetry analyses for the FM interlayer structures are shown in Table S2. Note that this is only valid for bilayer MSFEs (or even layer numbered MSFEs), while the trilayer (or odd layer numbered) MSFEs are different (see Figure S3 and Table S3 in SI).

|  | $D_\uparrow$ and $D_\downarrow$ | | IM | |
|---|---|---|---|---|
|  | Symmetry | Allowed photocurrents | Symmetry | Allowed photocurrents |
| **L** ∥ $\hat{\mathbf{x}}$ | $\mathcal{M}_x$ | $\eta^{yyy}, \eta^{yxx}$ | $\mathcal{M}_x, \mathcal{M}_z$ | $\eta^{yyy}, \eta^{yxx}$ |
| **L** ∥ $\hat{\mathbf{y}}$ | $\mathcal{M}_x \mathcal{T}$ | $\eta^{xxx}, \eta^{xyy}$ | $\mathcal{M}_x \mathcal{T}, \mathcal{M}_z$ | $\eta^{xxx}, \eta^{xyy}$ |
| **L** ∥ $\hat{\mathbf{z}}$ | $\mathcal{M}_x \mathcal{T}$ | $\eta^{xxx}, \eta^{xyy}$ | $\mathcal{M}_x \mathcal{T}, \mathcal{M}_z \mathcal{T}$ | all forbidden |

We show that $D_z$ can be used to control the direction and magnitude of MIC for the $\mathbf{L} \parallel \hat{\mathbf{z}}$ configuration. This can be understood by noting that the $D_\uparrow$ and $D_\downarrow$ states are connected via $\mathcal{M}_z \mathcal{T} D_\uparrow = D_\downarrow$. Thus, the transformation between $\eta_{D_\uparrow}^{xxx}$ and $\eta_{D_\downarrow}^{xxx}$ (for $\mathbf{L} \parallel \hat{\mathbf{z}}$) satisfies

$$\eta_{D_\downarrow}^{xxx} = \mathcal{M}_z \mathcal{T} \eta_{D_\uparrow}^{xxx} \sim \mathcal{M}_z \mathcal{T} \sum_{mn,\mathbf{k}} f_{nm}(\mathbf{k}) \Delta_{nm}^x(\mathbf{k}) g_{nm}^{xx}(\mathbf{k}) \delta(\omega_{mn}(\mathbf{k}) - \omega)$$

$$= -\sum_{mn,-\mathbf{k}} f_{nm}(-\mathbf{k}) \Delta_{nm}^x(-\mathbf{k}) g_{nm}^{xx}(-\mathbf{k}) \delta(\omega_{mn}(-\mathbf{k}) - \omega)$$

$$= -\sum_{mn,\mathbf{k}} f_{nm}(\mathbf{k}) \Delta_{nm}^x(\mathbf{k}) g_{nm}^{xx}(\mathbf{k}) \delta(\omega_{mn}(\mathbf{k}) - \omega) \sim -\eta_{D_\uparrow}^{xxx} = \eta_{D_\downarrow}^{xxx}. \qquad (4)$$

Here, the product of $g_{nm}^{bb}(\mathbf{k}) \delta(\omega_{mn} - \omega)$ measures the light absorption rate of each $\mathbf{k}$ point. This suggests that the MIC direction can be well-controlled by $D_z$ in the MSFE bilayer, which is significantly different from the nonmagnetic SFE bilayers[29] that vertical dipole flip keeps the photocurrent unchanged.

Similar behavior can be found in the $\mathbf{M} \parallel \hat{\mathbf{x}}$ and $\mathbf{M} \parallel \hat{\mathbf{y}}$ interlayer configurations of FM interlayer configuration (**M** denotes the spin polarization vector), as they share the same structural symmetry transformation rules with that of $\mathbf{L} \parallel \hat{\mathbf{z}}$. While for other magnetic configurations ($\mathbf{L} \parallel \hat{\mathbf{x}}$, $\mathbf{L} \parallel \hat{\mathbf{y}}$, and $\mathbf{M} \parallel \hat{\mathbf{z}}$), the $D_\uparrow$ and $D_\downarrow$ states are related to each other by $\mathcal{M}_z$. Hence, the MIC is unchanged under $D_z$ reversal. In this way, the MIC provides a facile way to detect the electric and magnetic configurations in MSFEs.

*First-principles calculation results.* In order to verify the above symmetry conclusions and quantify the photoconductivity, we perform first-principles DFT calculations. First of all, our magnetocrystalline anisotropy energy (MAE) calculation including full SOC effect shows that the magnetic moment prefers lying in the *xy* plane, while the $\mathbf{L} \parallel \hat{\mathbf{z}}$ is ~2 meV/f.u. higher in energy. This indicates that VSe$_2$ monolayer belongs to *XY* magnetic model, which could exhibit Berezinskii–Kosterlitz–Thouless (BKT) transition. This infers that the experimental measurements should be performed under low temperature, as in the observation of the CrCl$_3$ monolayer.[71] The in-plane MAE can be broken when one resorts to uniaxial strains, or carefully selecting specific substrate with strong anisotropy. The detailed results are shown in Figure S4, Tables S4 and S5. In addition, previous works have suggested that an external electric field can induce magnetoelectric Edelstein effect,[72-74] which could generate a sufficiently large effective magnetic field. In this way, the system could belong to the Ising model with fixed magnetic easy axis, aiding the potential experimental verifications of long-range magnetism.

The NSC is unaffected by $\mathcal{T}$ and $D_z$, which transports along *y* in all these cases (Figure 2, Figures S5 and S6). These results totally agree with the previous symmetry analyses. In the $\mathbf{L} \parallel \hat{\mathbf{z}}$ case, the $\mathcal{M}_x$ and $\mathcal{C}_3$ rotation symmetries lead to one independent in-plane non-vanishing shift current tensor, $\sigma^{yyy} = -\sigma^{yxx}$. From Figure 2, we see that finite NSC is generated across a large energy range due to the interband optical excitation above the bandgap. Thus, above bandgap open-circuit voltage is generated. The maximum value of $\sigma^{yyy}$ reaches 28 $\mu A/V^2$ at an incident light energy of 1.6 eV. Compared with previously proposed 2H-MoS$_2$ monolayer (photoconductivity 8 $\mu A/V^2$ at the photon energy of 2.8 eV),[75] the $\sigma^{yyy}$ in MSFE bilayer VSe$_2$ is much larger.

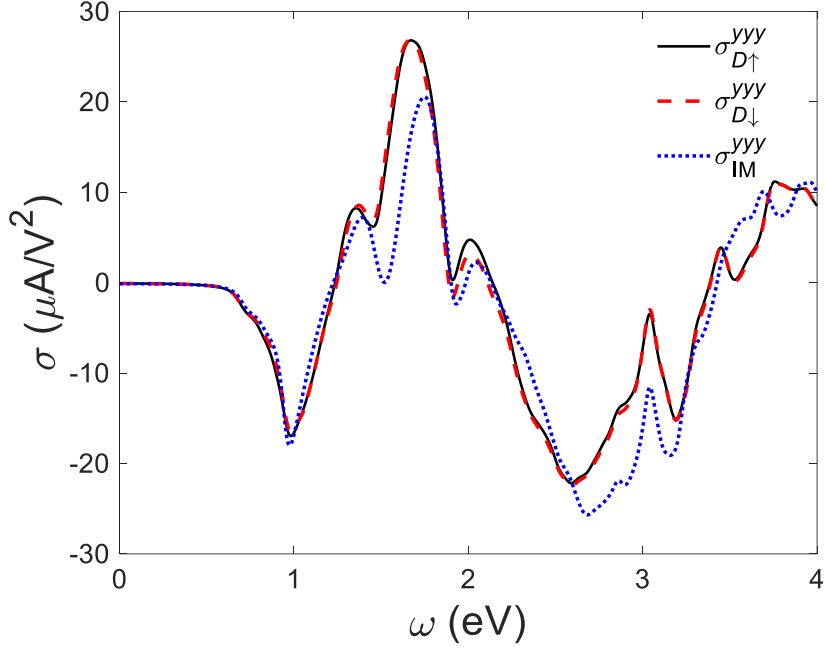

**Figure 2.** Incident light photon energy dependent NSC photoconductivities in the **L** ∥ **ẑ** IM, $D_\uparrow$, and $D_\downarrow$ states of bilayer VSe$_2$. One sees that flipping the dipole keeps the NSC.

Figures 3a and 3b depict the MIC photoconductivity of $\eta_{IM}$, $\eta_{D_\uparrow}$, and $\eta_{D_\downarrow}$ for the **L** ∥ **x̂**, and those for **L** ∥ **ŷ** are shown in Figures 3c and 3d. Consistent with previous symmetry results, for the **L** ∥ **x̂**, only the *y*-direction MIC survives. Rotating the magnetic axis to *y* (**L** ∥ **ŷ**) switches the MIC direction to *x*. This demonstrates that magnetic order **L** strongly affects the MIC feature. It should be noted that the **L** ∥ **ŷ** magnetization yields vertically propagating NSC and MIC, which could be detected and measured separately. In both the **L** ∥ **x̂** and **L** ∥ **ŷ** cases, flipping $D_z$ maintains the MIC. The magnitude of $\eta$ reaches 10 $\mu A/V^2$, when we assume the carrier lifetime $\tau$ to be 0.1 ps. This carrier lifetime accounts for the scattering that arises from the environment such as temperature, disorder, impurities, etc. Note that even though the carrier lifetime essentially depends on band index and momentum, a rigorous evaluation is not straightforward. Hence, we follow the conventional approach to adopt a universal value.[11] Compared to experimental measurements and theoretical estimations, e.g., 0.4 ps for CrI$_3$[32] and 0.2 ps for Ge,[76-77] our choice of 0.1 ps is conservative and may not overestimate the injection current magnitude. The MIC photoconductivity value indicates that under a LPL irradiation with its electric field magnitude of 3 V/nm, one could generate a current density of 90 μA/nm². Note that here we assume the effective thickness of VSe$_2$ bilayer to be 0.96 nm, which is estimated from its corresponding bulk structure.

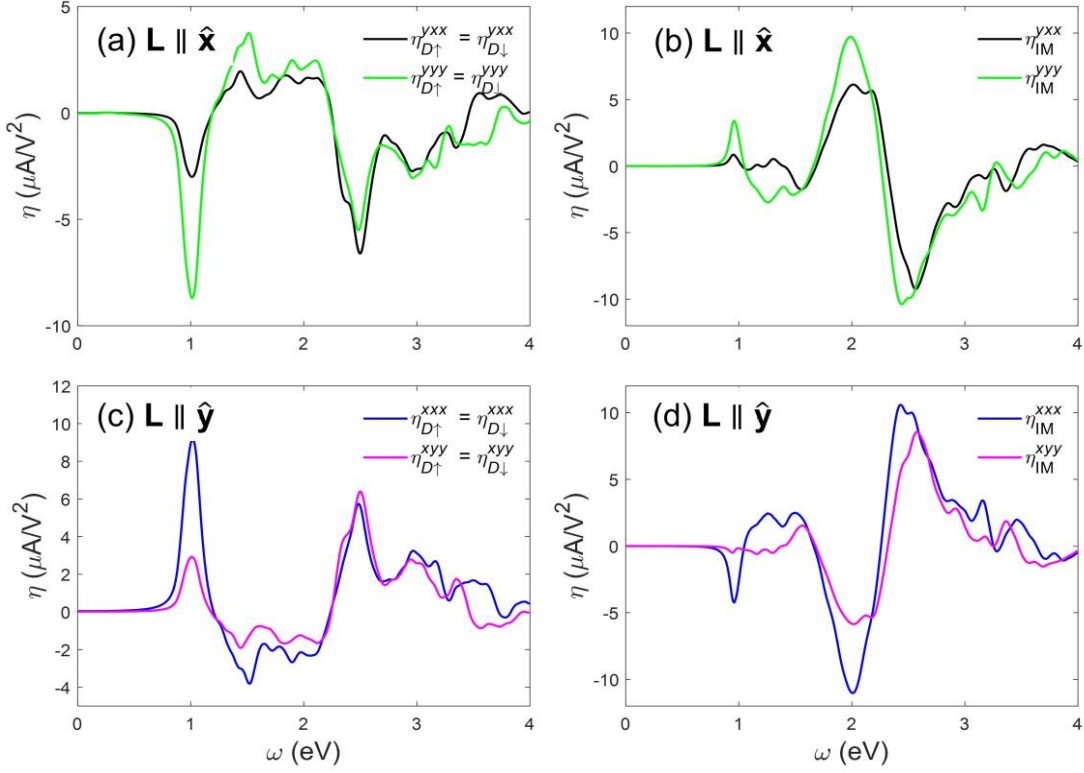

**Figure 3.** MIC for (a), (b) **L** ∥ $\hat{\mathbf{x}}$ and (c), (d) **L** ∥ $\hat{\mathbf{y}}$ of bilayer VSe$_2$. Both *x*- and *y*-LPL results are shown.

When the magnetization is along the *z*-axis, the MIC flowing along *y* is symmetrically forbidden. In this case, the system preserves $\mathcal{C}_3$ rotation. Hence, the *x* and *y*-LPL generate opposite MIC (Figure 4a), namely, $\eta^{xxx} = -\eta^{xyy}$ for both the $D_\uparrow$ and $D_\downarrow$ states. The MIC magnitude reaches ~2.8 μÅ/V² at the incident photon energy of ~1 eV (τ = 0.1 ps). In addition, flipping the MSFE dipole reverses the MIC, $\eta_{D\uparrow}^{xbb} = -\eta_{D\downarrow}^{xbb}$ (*b* = *x* or *y*). In the IM pattern, the MIC is symmetrically forbidden. These are well-consistent with our previous symmetry arguments. The frequency-dependent MIC photoconductivity responses for the FM configurations are shown in Figure S7. All these results demonstrate that the direction of the MIC can be controlled by **L** and $D_z$.

To further enlighten the absence and presence of the MIC and its $D_z$ control in the **L** ∥ $\hat{\mathbf{z}}$ bilayer VSe$_2$, we plot the BZ contribution of $\eta^{xxx}$ with the incident photon energy of 1 eV (Figures 4b−4d). One sees that the main contributions are from the vicinity of the *M* point, which arises from the electron transition between the Se-*p* (VB−3) and V-*d* (CB) orbitals (Figure 1e). The **k**-resolved $\Delta_{nm}^x$, $g_{nm}^{bb}$, and $\Delta_{nm}^x g_{nm}^{xx}$ distributions in the BZ for IM, $D_\uparrow$ and $D_\downarrow$ states are presented in Figure S8 and Figures 4b-d. For the IM state, it is evident that equally positive and negative proportion of the quantum metric dipole in BZ, $\Delta_{nm}^x(k_x,k_y)g_{nm}^{xx}(k_x,k_y) = -\Delta_{nm}^x(-k_x,-k_y)g_{nm}^{xx}(-k_x,-k_y)$, leading to cancellation of the $\eta^{xxx}$.[32] However, for the $D_\uparrow$ and $D_\downarrow$, one sees that the quantum metric dipole distributes inequivalently under $\mathbf{k} \to -\mathbf{k}$. For example, the positive quantum metric dipole at $(-k_x,-k_y)$ around $M$ surpasses that of the negative one at $(k_x,k_y)$ for $D_\uparrow$. This feature is reversed for $D_\downarrow$. Hence, the asymmetric quantum metric dipole in the BZ generates nonvanishing current. In addition, the opposite quantum metric dipole distributions for $D_\uparrow$ and $D_\downarrow$ reverse the sign of $\eta^{xxx}$.

One should note that the amount of $\mathbf{k}$ to $-\mathbf{k}$ symmetry breaking reflects the intensity of the MIC. According to our low energy model, the synergistic effects of $H'_D$ and $H'_{SOC}$ lead to valley polarization and asymmetric wavefunction distribution between $\mathbf{k}$ and $-\mathbf{k}$. Hence, both the parameters $\lambda_z$ and $\lambda_{SOC}$ control the MIC magnitude. In order to yield enhanced MIC photoconductivity, one could resort to materials that possess large sliding dipole moments with heavy elements.

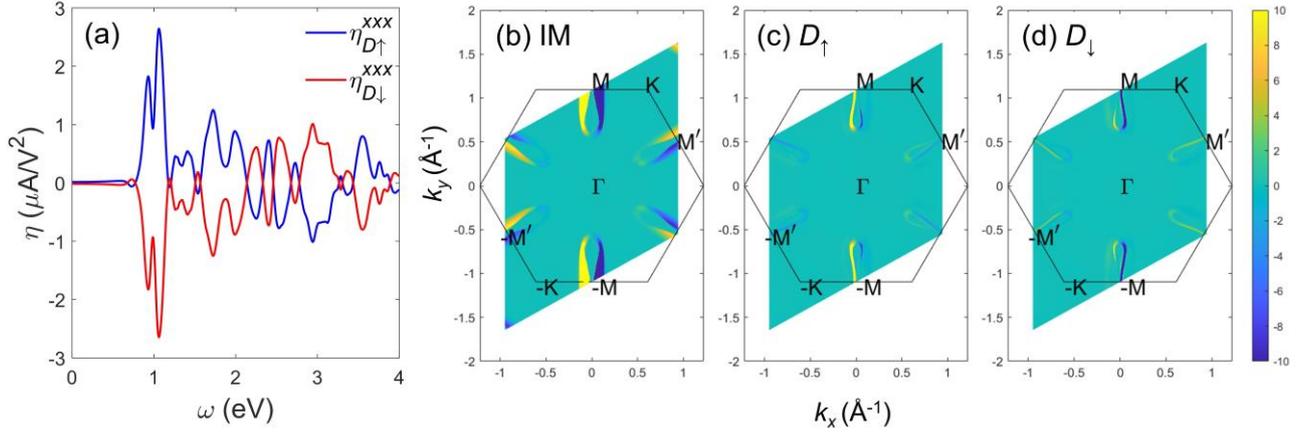

**Figure 4.** (a) MIC for $D_\uparrow$ and $D_\downarrow$ $\mathbf{L} \parallel \hat{\mathbf{z}}$ bilayer VSe$_2$ under $x$-LPL. (b-d) The quantum metric dipole ($\Delta_{nm}^x g_{nm}^{xx}$) ridges in the first BZ, which mainly corresponds to the electronic transition between the VB−3 and the lowest CB for (b) IM, (c) $D_\uparrow$, and (d) $D_\downarrow$ bilayer VSe$_2$. The main contribution to the $\eta^{xxx}$ is around $M$, under an incident photon energy of 1 eV.

Both the magnetic and electric dipole controlled BPV currents are also evidenced in the T-FeCl$_2$ and CrI$_3$ MSFE[78] systems. The relevant results are provided in Figures S9–S11 and S12, respectively. The out-of-plane easy axis of CrI$_3$ guarantees its belonging to the magnetic Ising model. In addition, it should be noted that our results are distinct from the $\mathcal{PT}$-symmetric cases in e.g., Ref. 35, where they showed that the MIC of MnBi$_2$Te$_4$ can be tuned by $\mathcal{T}$ and the NSC is switched by the out-of-plane electric field. Similarly, it has been demonstrated that shift current in twisted double bilayer graphene can be switched by the out-of-plane gate voltage which breaks the $\mathcal{C}_{2x}$ symmetry.[79]

In summary, we investigate the LPL illuminated NLO responses in the hexagonal MSFE

bilayers. Our symmetric analyses suggest that the NSC is unaffected by spin polarization, while the MIC photoconductivity is sensitive to the magnetic orientations. In this way, both magnetic and electric dipole could be utilized to control the BPV effect. Our first-principles results confirm these symmetric analyses on two different types of MSFE, bilayers H-VSe$_2$ and T-FeCl$_2$. The $D_z$ and SOC controlled valley polarization is captured by a simple **k·p** model, which elucidates the fundamental mechanism of nonvanishing MIC photoconductivity in MSFEs. Considering the profusion of 2D materials with the honeycomb lattice, magnetic configuration dependent photocurrent could find their potential applications in both optospintronic and optoelectronic devices.

**Methods.**

The DFT calculations on geometric and electronic calculations are performed within the Vienna *ab initio* simulation package (VASP)[80-82], using the projector augmented-wave (PAW) method[83] to treat the core electrons. A planewave basis set with a kinetic cutoff energy of 500 eV is used to describe the valence electrons. Generalized gradient approximation (GGA) in the form of Perdew-Burke-Ernzerhof (PBE) is applied to treat the exchange-correlation functional,[84] and the Hubbard *U* correction is adopted to treat the strong correlation in the magnetic *d* orbitals. The effective Hubbard *U* parameter is chosen to be 1.2 eV for the V-*d* and 4 eV for the Fe-*d* orbital, which have been proven to give results well consistent with experimental observations. Spin-orbit coupling (SOC) is added self-consistently throughout the calculations. A vacuum region of over 15 Å along the *z* direction is adopted to eliminate the artificial interactions between different images. The BZ integration is sampled by using Γ-centered Monkhorst−Pack **k**-point meshes with grid of (11×11×1). The vdW interactions are semi-empirically described according to the DFT-D3 method.[85] We use Wannier90 code[86-87] to construct the tight-binding model based on the maximally localized Wannier functions (MLWFs). The photo-conductivities are then integrated on a refined **k** grid of 500 × 500 × 1, which has been carefully tested to generate well-converged results.

The NSC [$\sigma^{abb}(0;\omega,-\omega)$] and MIC [$\eta^{abb}(0;\omega,-\omega)$] photo-conductivities can be evaluated according to band theory. Here the superscript *a* refers to current flow direction, and *b* is the LPL polarization direction. In the length-gauge formulism, they are

$$\sigma^{abb}(0;\omega,-\omega) = \frac{\pi e^3}{\hbar^2} \int \frac{d^3\mathbf{k}}{(2\pi)^3} \sum_{m,n} f_{mn} R_{mn}^{a;b} |r_{mn}^b|^2 \delta(\omega_{mn} - \omega) \qquad (5)$$

$$\eta^{abb}(0;\omega,-\omega) = -\frac{\tau \pi e^3}{2\hbar^2} \int \frac{d^3\mathbf{k}}{(2\pi)^3} \sum_{m,n} f_{mn} \Delta_{mn}^a |r_{mn}^b|^2 \delta(\omega_{mn} - \omega) \qquad (6)$$

where the sum is over the band indices *m* and *n*. The integration is performed in the first BZ, which includes vacuum space contribution in the supercell. In order to correct this and keep the photocurrent unit consistent with conventional 3D bulk systems, we use the well-adopted scaling approach by multiply a factor $\frac{L_z}{d_{\text{eff}}}$ onto the supercell results. Here $L_z$ is the lattice constant along the z-axis (including vacuum), and $d_{\text{eff}}$ is the effective thickness of the system. $f_{mn} = f_{m\mathbf{k}} - f_{n\mathbf{k}}$ is the difference of Fermi-Dirac occupation between bands. The broadening factor of the Dirac delta function is taken to be 0.04 eV. The shift vector $R_{mn}^{a;b}$ is defined as

$$R_{mn}^{a;b} = \partial_a \phi_{mn}^b - A_{mm}^a + A_{nn}^a. \tag{7}$$

Here, $\phi_{mn}^b$ is the phase of $r_{mn}^b$ ($= |r_{mn}^b|e^{i\phi_{mn}^b}$), and $A_{mm}^a$ is the intraband Berry connection $A_{mm}^a = i\langle m|\partial_a m\rangle$. The $|r_{mn}^b|^2 \delta(\omega_{nm} - \omega)$ evaluates the absorption rate from band $m$ to band $n$, according to the Fermi's golden rule. $\Delta_{nm}^a = v_{nn}^a - v_{mm}^a$ is the group velocity difference between the $n$-th and the $m$-th bands. The explicit **k**-dependence on these quantities is omitted.

Under LPL, Eq. (6) is equivalent to the contribution from quantum metric tensor $g_{nm}^{bb} = \sum_{\mu,\nu} \text{Re}\left(r_{m_\nu n_\mu}^b r_{n_\mu m_\nu}^b\right)$, where $\mu, \nu$ represent the degenerate bands arising from the potential Kramers degeneracy (for example in AFM configurations),[34]

$$\eta^{a,bb}(0;\omega,-\omega) = -\frac{\tau\pi e^3}{2\hbar^2} \int \frac{d^3\mathbf{k}}{(2\pi)^3} f_{mn} \Delta_{mn}^a (2g_{mn}^{bb}) \delta(\omega_{mn} - \omega) \tag{8}$$

Accordingly, the MIC arises from an excitation between $m$ and $n$, where $g_{nm}^{bb}$ is weighted by the asymmetric group velocity difference $\Delta_{nm}^a$ at the time-reversed $\pm\mathbf{k}$ pairs. The $\Delta_{nm}^a g_{nm}^{bb}$ is defined as quantum metric dipole.[88]

**Associated content**

**Supporting Information:** The Supporting Information is available free of charge at https://pubs.acs.org/doi/..........

Energy comparison for different magnetic configurations; The explicit reason for the valley polarization in **L** ∥ **ẑ** $D_\uparrow$ and $D_\downarrow$ bilayer VSe$_2$; The symmetry analyses of the allowed photoconductivity for FM bilayer VSe$_2$; The asymmetric bands associated nonvanishing MIC; The angle dependence of total MAE for MSFE VSe$_2$; Switchable MAE in bilayer H-VSe$_2$; Frequency-dependent MIC responses of FM bilayer VSe$_2$; Quantum metric, velocity, and quantum metric dipole distributions in the first BZ; The NSC for all magnetic configurations; The photocurrent analyses for T-FeCl$_2$ MSFE; Photocurrent in trilayer cases; Switchable photocurrent in CrI$_3$ MSFE.

**Data availability**

All the data and methods are present in the main text and the Supporting Information. Any other relevant data are available from the authors upon reasonable request.

**Notes**

The authors declare no competing financial interest.

**Acknowledgement**

We acknowledge the financial support from the National Natural Science Foundation of China (NSFC) under Grant Nos. 12004306, 11974270, 21903063, 12274342, and 11974277.